\documentclass[]{acm_proc_article-sp}
\usepackage{url}
\usepackage[pdf]{pstricks}
\usepackage{pstricks-add, pst-grad, pst-plot}
\usepackage[tiling]{pst-fill}
\psset{linewidth=0.3pt,dimen=middle}
\psset{xunit=.70cm,yunit=0.70cm}
\psset{angleA=-90,angleB=90,ArrowInside=->,arrowscale=2}

\usepackage{bigpage}
\usepackage{bcprules}
\usepackage{mathpartir}
\usepackage{listings}
\usepackage{mathtools}
\usepackage{amsfonts}
\usepackage{latexsym}
\usepackage{amssymb}
\usepackage{caption}

\newcommand{\maps}{\colon}

\newcommand{\ldb}{[\![}
\newcommand{\rdb}{]\!]}

\newcommand{\id}[1]{\texttt{#1}}

\newcommand{\pzero}{\mathbin{0}}

\newcommand{\juxtap}{\mathbin{\id{|}}}
\newcommand{\concat}{\Rightarrow}

\newcommand{\scong}{\mathbin{\equiv}}

\newcommand{\alphaeq}{\mathbin{\equiv_{\alpha}}}
\newcommand{\names}[1]{\mathbin{\mathcal{N}(#1)}}
\newcommand{\freenames}[1]{\mathbin{\mathcal{FN}(#1)}}

\newcommand{\binpar}[2]{#1 \juxtap #2}

\newcommand{\prefix}[3]{#1 ? ( #2 ) \concat #3}

\newcommand{\meaningof}[1]{\ldb #1 \rdb}

\newcommand{\defneqls}{\coloneqq}

\renewcommand{\red}{\rightarrow}

\newcommand{\rel}[1]{\;{\mathcal #1}\;} 
\newcommand{\wbbisim}{\stackrel{\centerdot}{\approx}} 
\newcommand{\wcc}{\stackrel{\centerdot}{\simeq}}
\newcommand{\wbbisimsem}{\approx} 

\newcommand{\bc}{\mathbin{\mathbf{::=}}}
\newcommand{\bm}{\mathbin{\mathbf\mid}}

\newlength{\ltext}
\newlength{\lmath}
\newlength{\cmath}
\newlength{\rmath}
\newlength{\rtext}

\newenvironment{grammar}{
  \[
  \begin{array}{l@{\quad}rcl@{\quad}l}
  \hspace{\ltext} & \hspace{\lmath} & \hspace{\cmath} & \hspace{\rmath} & \hspace{\rtext} \\
}{
  \end{array}\]
}

 \newtheorem{thm}{Theorem}[subsection]
 
 \newtheorem{lem}[thm]{Lemma}

 \newtheorem{defn}[thm]{Definition}

 \numberwithin{equation}{subsection}

\newcommand{\pic}{$\pi$-calculus}

\newcommand{\papertitle}{Higher category models of the pi-calculus}

\title{\papertitle}

\author{
Michael Stay\\
  \affaddr{Google}\\
  \email{\fontsize{8}{8}\selectfont stay@google.com}
\and
L.G. Meredith\\
  \affaddr{Biosimilarity, LLC}\\
  \email{\fontsize{8}{8}\selectfont lgreg.meredith@biosimilarity.com}
}

\begin{document}
\lstset{language=}

\setlength{\topmargin}{0in}
\setlength{\textheight}{8.5in}
\setlength{\parskip}{6pt}


\begin{abstract}
\normalsize{ 

  We present an approach to modeling computational calculi using
  higher category theory. Specifically we present a fully abstract
  semantics for the {\pic}. The interpretation is consistent with
  Curry-Howard, interpreting terms as typed morphisms, while
  simultaneously providing an explicit interpretation of the rewrite
  rules of standard operational presentations as 2-morphisms. One of
  the key contributions, inspired by catalysis in chemical reactions,
  is a method of restricting the application of 2-morphisms
  interpreting rewrites to specific contexts.

}

\end{abstract}


\maketitle




\section{Introduction}

One of the major distinctions in programming language semantics has
been the division between denotational and operational semantics. In
the former computations are interpreted as mathematical objects
which---more often than not---completely unfold the computational
dynamics, and are thus infinitary in form. In the latter computations
are interpreted in terms of rewrite rules operating on finite
syntactic structure. Historically, categorical semantics for
programming languages, even variations such as games semantics
\cite{DBLP:conf/lics/McCusker96} which capture much more of the
intensional structure of computations, are distinctly denotational in
flavor \cite{DBLP:journals/iandc/Moggi91}. Meanwhile, operational
semantics continues to dominate in the presentation of calculi
underlying programming languages used in practice
\cite{DBLP:conf/aplas/MaffeisMT08}
\cite{DBLP:conf/oopsla/IgarashiPW99}
\cite{Politz:2013:PFM:2509136.2509536}.

Motivated, in part, by the desire to make a closer connection between
theory and practice, many efforts in the programming language
semantics, and in concurrency theory communities have begun to
investigate more direct categorical interpretations of operational
semantics. This paper finds its place in this latter context,
providing a fully abstract interpretation of the {\pic} in terms of a
higher categorical model of its operational semantics. In particular,
while it remains faithful to a Curry-Howard orientation, modeling
terms as typed morphisms, it models the computational dynamics of the
calculus, its rewrite rules, as 2-morphisms. One of the goals has been
to provide a modular semantics to address a range of features and
modeling options typically associated with the {\pic}. For example, a
significant bifurcation occurs in the treatment of names with Milner's
original calculus hiding all internal structure of names
\cite{milner91polyadicpi}, while the $\rho$-calculus variant provides
a reflective version in which names are the codes of processes
\cite{DBLP:journals/entcs/MeredithR05}. The semantics presented here
is capable of providing a categorical interpretation of both variants.

Of particular interest to theoreticians and implementers, the
semantics shines light on a key difference between the categorical and
computational machinery it interprets. The latter is intrinsically
lazy in the sense that all contexts where rewrites can apply must be
explicitly spelled out (cf the context rules in section
\ref{section:opsem}), while the former is intrinsically eager \footnote{like the mythical hydra, chop off one 1-morphism and a 2-morphism takes its place ;-)}; in
fact, one of the contributions of the paper is the delineation of an
explicit control mechanism to prevent unwanted rewrites that would
otherwise create an insurmountable divergence between the two
formalisms. 

\subsubsection{Related work}


This paper draws inspiration from \cite{DBLP:conf/lics/Seely87} and
\cite{DBLP:journals/tcs/Hilken96}, but also seeks a more direct
account of what works in modern day operational semantics. In his
seminal paper \cite{DBLP:journals/mscs/Milner92} Milner provided the
template still used today for specifying computational calculi,
presenting the {\pic} in terms of a freely generated algebra
quotiented by a structural equivalence relation that is then subject
to some rewrite rules. This constitutes the modern view of structured
operational semantics \cite{Plotkin04theorigins}. In the latter part
of his research Milner focused on finding a satisfying relationship
between a categorical presentation of the rewrite rules and the notion
of bisimulation \cite{DBLP:conf/concur/LeiferM00}. While this work did
not explicitly employ higher categorical techniques, it spawned a
variety of 2-categorical investigations designed to capture and recast
bisimulation equivalences in terms of 2-morphisms
\cite{Sassone02derivingbisimulation}. Hirschowitz has developed an even
more ambitious program of categorifying the whole of the operational
semantics framework from the presentation of higher order syntax (or
terms with binding constructors like {\pic} or $\lambda$-calculus), to
rewrite rules \cite{hirschowitzcc2c}.

The present work is primarily focused on providing a direct account of
the {\pic}. The modularity of the semantics arises from wanting to
give a clean design and clear shape to the present account, rather
than an attempt to provide a framework for interpreting a number of
computational calculi. The fact that the techniques do apply to a
number of calculi was a side effect of this process. Moreover, our
particular reconciliation of operational laziness with categorical
eagerness introduces an explicit resource sensitivity, which we have
not seen before in the theoretical literature, yet is remarkably
similar to resource constraints in actual implementations of
concurrent and distributed computations.

\subsubsection{Organization of the rest of the paper}

In the remainder of the paper we present the core fragment of the
calculus we model followed by a manifest of the categorical equipment
needed to faithfully model it. Then we give the semantics function an
sketch a proof that the interpretation is fully abstract.


\section{The calculus}

One notable feature of the {\pic} is its ability to succinctly and
faithfully model a number of phenomena of concurrent and distributed
computing. Competition for resources amongst autonomously executing
processes is a case in point. The expression
\begin{equation*}
  x?( y ) \Rightarrow P \juxtap x!( u ) \juxtap x?( v ) \Rightarrow Q
\end{equation*}
is made by composing three processes, two of which, $x?( y )
\Rightarrow P$ and $x?( v ) \Rightarrow Q$ are seeking input from
channel $x$ before they launch their respective continuations, $P$
and/or $Q$; while the third, $x!( u )$, is supplying output on that
same said channel. Only one of the input-guarded processes will win,
receiving $u$ and binding it to the input variable, $y$, or
respectively, $v$ in the body of the corresponding continuation --
while the loser remains in the input-guarded state awaiting input
along channel $x$. The calculus is equinanimous, treating both
outcomes as equally likely, and in this regard is unlike its
sequential counterpart, the $\lambda$-calculus, in that it is not
\emph{confluent}. There is no guarantee that the different branches of
computation must eventually converge. Note that just adding a
$\mathsf{new}$-scope around the expression
\begin{equation*}
  (\mathsf{new}\; x)( x?( y ) \Rightarrow P \juxtap x!( u ) \juxtap x?( v ) \Rightarrow Q )
\end{equation*}
ensures that the competition is for a local resource, hidden from any
external observer.

\subsection{Our running process calculus}

\subsubsection{Syntax}
\label{syntax}
\begin{grammar}
{P} \bc \pzero & \mbox{stopped process} \\
       \;\;\; \bm \; {x}{?}{( y_1, \ldots, y_n )} \Rightarrow {P} & \mbox{input} \\
       \;\;\; \bm \; {x}{!}{( y_1, \ldots, y_n )} & \mbox{output} \\
       \;\;\; \bm \; (\mathsf{new}\; x){P} & \mbox{new channel} \\
       \;\;\; \bm \; {P} \juxtap {Q} & \mbox{parallel} \\                                
\end{grammar}

Due to space limitations we do not treat replication, $!P$.

\subsubsection{Free and bound names}

\begin{equation*}
  \begin{aligned}
    & \freenames{\pzero} \defneqls \emptyset \\
    & \freenames{{x}{?}{( y_1, \ldots, y_n )} \Rightarrow {P}} \defneqls \\
    & \;\;\;\;\;\{ x \} \cup (\freenames{P} \setminus \{ y_1, \ldots y_n \}) \\
    & \freenames{{x}{!}{( y_1, \ldots, y_n )}} \defneqls \{ x, y_1, \ldots, y_n \} \\
    & \freenames{(\mathsf{new}\; x){P}} \defneqls \freenames{P} \setminus \{x\} \\    
    & \freenames{{P} \juxtap {Q}} \defneqls \freenames{P} \cup \freenames{Q} \\
  \end{aligned}
\end{equation*}

An occurrence of $x$ in a process $P$ is \textit{bound} if it is not
free. The set of names occurring in a process (bound or free) is
denoted by $\names{P}$.

\subsubsection{Structural congruence}
\label{congruence}

The {\em structural congruence} of processes, noted $\scong$, is the
least congruence containing $\alpha$-equivalence, $\alphaeq$, making
$( P, |, 0 )$ into commutative monoids and satisfying

\begin{equation*}  
  (\mathsf{new}\; x)(\mathsf{new}\; x){P} \scong (\mathsf{new}\; x)P
\end{equation*}
\begin{equation*}  
  (\mathsf{new}\; x)(\mathsf{new}\; y){P} \scong (\mathsf{new}\; y)(\mathsf{new}\; x)P
\end{equation*}
\begin{equation*}  
  ((\mathsf{new}\; x){P}) \juxtap {Q} \scong (\mathsf{new}\; x)({P} \juxtap {Q})
\end{equation*}

\subsubsection{Operational Semantics}\label{section:opsem}
 
\infrule[Comm]
{ |\vec{y}| = |\vec{z}| }
{{{ x{?}{(}{\vec{y}}{)} \concat {P}}\juxtap {x}{!}{(}{\vec{z}}{)}}
\red {{P}{\{}\vec{z}{/}{\vec{y}}{\}}}}

In addition, we have the following context rules:

\infrule[Par]{{P} \red {P}'}{{{P} \juxtap {Q}} \red {{P}' \juxtap {Q}}}

\infrule[New]{{P} \red {P}'}{{(\mathsf{new}\; x){P}} \red {(\mathsf{new}\; x){P}'}}

\infrule[Equiv]{{{P} \scong {P}'} \andalso {{P}' \red {Q}'} \andalso {{Q}' \scong {Q}}}{{P} \red {Q}}

\subsubsection{Bisimulation}

\begin{defn}
An \emph{observation relation}, $\downarrow$ is the smallest relation satisfying the rules
below.

\infrule[Out-barb]{ }
      {{x}!(\vec{y}) \downarrow x}
\infrule[Par-barb]{\mbox{$P\downarrow x$ or $Q\downarrow x$}}
      {\mbox{$P \juxtap Q \downarrow x$}}
\infrule[New-barb]{\mbox{$P\downarrow x$, $x \neq u$}}
      {\mbox{$(\mathsf{new}\; u){P} \downarrow x$}}

\end{defn}

Notice that $\prefix{x}{y}{P}$ has no barb.  Indeed, in {\pic} as well
as other asynchronous calculi, an observer has no direct means to
detect if a sent message has been received or not.

\begin{defn}
An \emph{barbed bisimulation}, is a symmetric binary relation 
${\mathcal S}$ between agents such that $P\rel{S}Q$ implies:
\begin{enumerate}
\item If $P \red P'$ then $Q \red Q'$ and $P'\rel{S} Q'$.
\item If $P\downarrow x$, then $Q\downarrow x$.
\end{enumerate}
$P$ is barbed bisimilar to $Q$, written
$P \wbbisim Q$, if $P \rel{S} Q$ for some barbed bisimulation ${\mathcal S}$.
\end{defn}

\section{Categorical machinery}

We take our models in 2-categories with an underlying symmetric monoidal closed category; the 2-categories Cat (categories, functors, and natural transformations) and Rel (sets, relations, and implications) are examples.  We denote the monoidal unit object by $I$, the tensor product by $\otimes,$ the $n$th tensor power of an object $X$ by $X^{\otimes n},$ and the internal hom by a lollipop $\multimap$.

\section{The interpretation}

Given the abstract syntax of a term calculus like that in section \ref{syntax}, we introduce an object in our 2-category for each parameter of the calculus.  We introduce 1-morphisms for each term constructor, 2-morphisms for each reduction relation, and equations for structural equivalence; we also add 1-morphisms to mark contexts in which reductions may occur.

The {\pic} is parametric in a set of names and a set of processes, so we have objects $\mathcal{N}$ and $\mathcal{P}$.  Since names can be reused in the {\pic,} we also add 1-morphisms and equations to make $\mathcal{N}$ be a cocommutative comonoid.  We denote comultiplication by $\Delta\maps \mathcal{N} \to \mathcal{N} \otimes \mathcal{N}$ and counit by $\delta\maps \mathcal{N} \to I.$  If the tensor product is the cartesian product, $I$ is the terminal object, $\mathcal{N}$ is a comonoid in a unique way, and $\Delta$ and $\delta$ are duplication and deletion, respectively.

In the {\pic,} all reductions occur at the topmost context, so we have one unary morphism from $\mathcal{P}$ to $\mathcal{P}$.  There are some benefits to constructing the top context marker out of the existing binary morphism $|\maps \mathcal{P} \otimes \mathcal{P} \to \mathcal{P}$ and a unary morphism $COMM\maps I \to \mathcal{P};$ we'll talk about some of the benefits in the conclusion.

The theory of the {\pic} is the free symmetric monoidal closed 2-category on
\begin{itemize}
  \item objects $\mathcal{N}$ for names and $\mathcal{P}$ for processes,
    
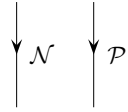
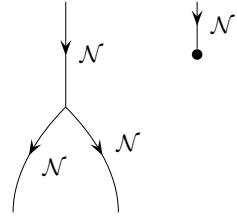
\[\begin{pspicture}(-.5,0)(0.5,2)
      \pnode(0,2){A}
      \pnode(0,0){B}
      \nccurve{A}{B} \naput{$\mathcal{N}$}
    \end{pspicture}\quad
    \begin{pspicture}(-.5,0)(0.5,2)
      \pnode(0,2){A}
      \pnode(0,0){B}
      \nccurve{A}{B} \naput{$\mathcal{P}$}
    \end{pspicture}\]
  \item 1-morphisms $\Delta\maps \mathcal{N} \to \mathcal{N} \otimes \mathcal{N}$ and $\delta\maps \mathcal{N} \to I,$
    \[\begin{pspicture}(0,0)(3,4)
      \pnode(1,4){A}
      \pnode(1,2){B}
      \nccurve{A}{B} \naput{$\mathcal{N}$}
      \pnode(0,0){C}
      \pnode(2,0){D}
      \nccurve[angleA=225]{B}{C} \naput{$\mathcal{N}$}
      \nccurve[angleA=315]{B}{D} \naput{$\mathcal{N}$}
    \end{pspicture}
    \begin{pspicture}(-.5,-3)(0.5,1)
      \rput(0,0){\cnode*{2pt}{A}}
      \pnode(0,1){B}
      \nccurve{B}{A} \naput{$\mathcal{N}$}
    \end{pspicture}\]
  \item 1-morphisms $|\maps \mathcal{P} \otimes \mathcal{P} \to \mathcal{P}$ and $0\maps I \to \mathcal{P},$
    \[\begin{pspicture}(0,0)(3,4)
      \rput(1,2){\ovalnode{A}{$|$}}
      \pnode(1,0){B}
      \nccurve{A}{B} \naput{$\mathcal{P}$}
      \pnode(0,4){C}
      \nccurve[angleB=135]{C}{A} \naput{$\mathcal{P}$}
      \pnode(2,4){D}
      \nccurve[angleB=45]{D}{A} \naput{$\mathcal{P}$}
    \end{pspicture}
    \begin{pspicture}(-.5,0)(0.5,4)
      \rput(0,2){\ovalnode{A}{$0$}}
      \pnode(0,0){B}
      \nccurve{A}{B} \naput{$\mathcal{P}$}
    \end{pspicture}\]
  \item 1-morphism $?_n\maps \mathcal{N} \otimes (\mathcal{N}^{\otimes n} \multimap \mathcal{P}) \to \mathcal{P}$ and $!_n\maps \mathcal{N} \otimes \mathcal{N}^{\otimes n} \to \mathcal{P}$ for each natural number $n \ge 0,$
    
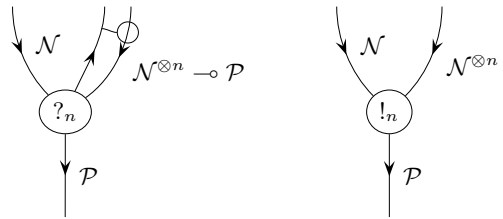
\[\begin{pspicture}(0,0)(6,4)
      \rput(1,2){\ovalnode{A}{$?_n$}}
      \pnode(1,0){B}
      \nccurve{A}{B} \naput{$\mathcal{P}$}
      \pnode(0,4){C}
      \nccurve[angleB=135]{C}{A} \naput{$\mathcal{N}$}
      \pnode(1.75,4){D}
      \nccurve[angleA=65,angleB=-90]{A}{D} \ncput[npos=.75]{\pnode{clasp1}}
      \pnode(2.25,4){E}
      \nccurve[angleB=35]{E}{A} \naput{$\mathcal{N}^{\otimes n}\multimap \mathcal{P}$} \ncput[npos=.25]{\cnode{4pt}{clasp2}}
      \nccurve[angleA=0,angleB=180,ArrowInside=]{clasp1}{clasp2}
    \end{pspicture}
    \begin{pspicture}(0,0)(2,4)
      \rput(1,2){\ovalnode{A}{$!_n$}}
      \pnode(1,0){B}
      \nccurve{A}{B} \naput{$\mathcal{P}$}
      \pnode(0,4){C}
      \nccurve[angleB=135]{C}{A} \naput{$\mathcal{N}$}
      \pnode(2,4){D}
      \nccurve[angleB=45]{D}{A} \naput{$\mathcal{N}^{\otimes n}$}
    \end{pspicture}\]
  \item 1-morphisms $fresh\maps I \to \mathcal{P}$ and $COMM\maps I \to \mathcal{P}$
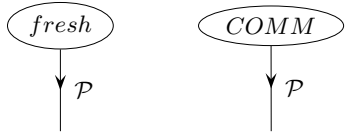
    \[\begin{pspicture}(-.5,0)(3.5,2.5)
      \rput(0,2){\ovalnode{A}{$fresh$}}
      \pnode(0,0){B}
      \nccurve{A}{B} \naput{$\mathcal{P}$}
    \end{pspicture}
    \begin{pspicture}(-.5,0)(0.5,2.5)
      \rput(0,2){\ovalnode{A}{$COMM$}}
      \pnode(0,0){B}
      \nccurve{A}{B} \naput{$\mathcal{P}$}
    \end{pspicture}\]
    (Note that the equations governing $\mathsf{new}$ in section \ref{congruence} are satisfied up to tensoring with a scalar due to the naturality of the unitors and braiding in the symmetric monoidal 2-category.)
    For convenience we write $\meaningof{x}$ for
     
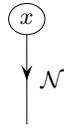
\[\begin{pspicture}(-.5,0)(3.5,2.5)
      \rput(0,2){\ovalnode{A}{$x$}}
      \pnode(0,0){B}
      \nccurve{A}{B} \naput{$\mathcal{N}$}
    \end{pspicture}\]
    which picks out $x$ from $\mathcal{N}$.
  \item equations making $(\mathcal{P}, |, 0)$ into a commutative monoid,
  \item equations making $(N, \Delta, \delta)$ into a cocommutative comonoid, and
  \item a 2-morphism $comm_n$ encoding the COMM rule for each natural number $n \ge 0.$
    \[\begin{pspicture}(-.5,0)(10,8)
      \pnode(0,8){A}
      \pnode(0,7){B}
      \pnode(1,8){C}
      \rput(1,5){\ovalnode{D}{$!_n$}}
      \nccurve{A}{B} \nbput{$\mathcal{N}$} 
      \nccurve[angleA=225,angleB=135]{B}{D}
      \nccurve[angleB=45]{C}{D} \naput[npos=.25]{$\mathcal{N}^{\otimes n}$}
      \pnode(5,8){E}
      \pnode(6,8){F}
      \rput(5,5){\ovalnode{G}{$?_n$}}
      \nccurve[angleA=315,angleB=135]{B}{G}
      \nccurve[angleA=90,angleB=270]{G}{E} \ncput[npos=.75]{\pnode{clasp1}}
      \nccurve[angleB=45]{F}{G} \ncput[npos=.25]{\cnode{4pt}{clasp2}} \naput{$\mathcal{N}^{\otimes n} \multimap \mathcal{P}$}
      \nccurve[angleA=0,angleB=180,ArrowInside=]{clasp1}{clasp2}
      \rput(8,5){\ovalnode{H}{$COMM$}}
      \rput(2.5,3){\ovalnode{I}{$|$}}
      \rput(3.5,1.5){\ovalnode{J}{$|$}}
      \pnode(3.5,0){K}
      \nccurve[angleB=135]{D}{I} \nbput{$\mathcal{P}$}
      \nccurve[angleB=45]{G}{I} \naput{$\mathcal{P}$}
      \nccurve[angleB=135]{I}{J} \nbput{$\mathcal{P}$}
      \nccurve[angleB=45]{H}{J} \naput{$\mathcal{P}$}
      \nccurve{J}{K} \nbput{$\mathcal{P}$}
    \end{pspicture}\]
    \[comm_n\Downarrow\]
    \[\begin{pspicture}(-.5,0)(10,8)
      \pnode(0,8){A}
      \rput(0,7){\cnode*{2pt}{B}}
      \pnode(1,8){C}
      \nccurve{A}{B} \nbput{$\mathcal{N}$} 
      \rput(2.5,5){\cnode{18pt}{D}}
      \nccurve[angleB=135]{C}{D} \naput{$\mathcal{N}^{\otimes n}$} \ncput[npos=1]{\pnode{Z}}
      \pnode(5,8){E}
      \pnode(6,8){F}
      \nccurve[angleA=90,angleB=270]{D}{E} \ncput[npos=.75]{\pnode{clasp1}} \ncput[npos=0]{\pnode{Y}}
      \nccurve[angleB=45]{F}{D} \ncput[npos=.25]{\cnode{4pt}{clasp2}} \naput{$\mathcal{N}^{\otimes n} \multimap \mathcal{P}$} \ncput[npos=1]{\pnode{X}}
      \nccurve[angleA=0,angleB=180,ArrowInside=]{clasp1}{clasp2}
      \rput(8,5){\ovalnode{H}{$COMM$}}
      \rput(3.5,1.5){\ovalnode{J}{$|$}}
      \pnode(3.5,0){K}
      \nccurve[angleB=135]{D}{J} \nbput{$\mathcal{P}$} \ncput[npos=0]{\pnode{W}}
      \nccurve[angleB=45]{H}{J} \naput{$\mathcal{P}$}
      \nccurve{J}{K} \nbput{$\mathcal{P}$}
      \nccurve[angleA=315,angleB=270]{Z}{Y}
      \nccurve[angleA=225,angleB=90]{X}{W}
      \rput(2,5){$ev$}
    \end{pspicture}\]
\end{itemize}
\subsection{Semantics}
\begin{description}
  \item $\meaningof{Q}_{top} \defneqls$
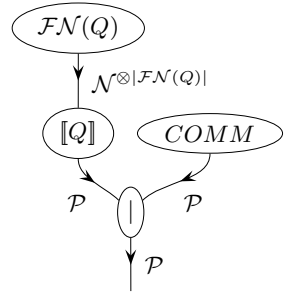
    \[\begin{pspicture}(0,0)(3,5.5)
      \rput(0,5){\ovalnode{Z}{$\freenames{Q}$}}
      \rput(0,3){\ovalnode{A}{$\meaningof{Q}$}}
      \rput(2.5,3){\ovalnode{B}{$COMM$}}
      \rput(1,1.5){\ovalnode{C}{$|$}}
      \pnode(1,0){D}
      \nccurve[angleB=135]{A}{C} \nbput{$\mathcal{P}$}
      \nccurve[angleB=45]{B}{C} \naput{$\mathcal{P}$}
      \nccurve{C}{D} \naput{$\mathcal{P}$}
      \nccurve{Z}{A} \naput{$\mathcal{N}^{\otimes |\freenames{Q}|}$}
    \end{pspicture}\]
  \item $\meaningof{\pzero} \defneqls$
  
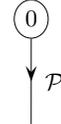
\[\begin{pspicture}(0,0)(0,2)
    \rput(0,2){\ovalnode{A}{$0$}}
    \pnode(0,0){B}
    \nccurve{A}{B} \naput{$\mathcal{P}$}
  \end{pspicture}\]
  \item $\meaningof{{x}{?}{( y_1, \ldots, y_n )} \Rightarrow {Q}} \defneqls$
    
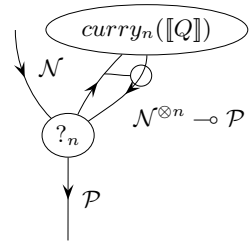
\[\begin{pspicture}(0,0)(0,4.5)
      \rput(1,2){\ovalnode{A}{$?_n$}}
      \pnode(1,0){B}
      \nccurve{A}{B} \naput{$\mathcal{P}$}
      \pnode(0,4){C}
      \nccurve[angleB=135]{C}{A} \naput{$\mathcal{N}$}
      \rput(2.5,4){\ovalnode{D}{$curry_n(\meaningof{Q})$}}
      \nccurve[angleA=65,angleB=-135]{A}{D} \ncput[npos=.65]{\pnode{clasp1}}
      \nccurve[angleA=-90,angleB=35]{D}{A} \naput{$\mathcal{N}^{\otimes n}\multimap \mathcal{P}$} \ncput[npos=.25]{\cnode{4pt}{clasp2}}
      \nccurve[angleA=0,angleB=180,ArrowInside=]{clasp1}{clasp2}
    \end{pspicture}\]
  \item $\meaningof{{x}{!}{( y_1, \ldots, y_n )}} \defneqls$
  
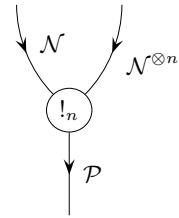
\[\begin{pspicture}(-1,0)(0,4.5)
    \rput(1,2){\ovalnode{A}{$!_n$}}
    \pnode(1,0){B}
    \nccurve{A}{B} \naput{$\mathcal{P}$}
    \pnode(0,4){C}
    \nccurve[angleB=135]{C}{A} \naput{$\mathcal{N}$}
    \pnode(2,4){D}
    \nccurve[angleB=45]{D}{A} \naput{$\mathcal{N}^{\otimes n}$}
  \end{pspicture}\]
  \item $\meaningof{(\mathsf{new}\;x)Q} \defneqls$
  
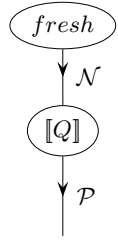
\[\begin{pspicture}(-.5,0)(.5,4)
    \rput(0,4){\ovalnode{A}{$fresh$}}
    \rput(0,2){\ovalnode{B}{$\meaningof{Q}$}}
    \pnode(0,0){C}
    \nccurve{A}{B} \naput{$\mathcal{N}$}
    \nccurve{B}{C} \naput{$\mathcal{P}$}
  \end{pspicture}\]
  \item $\meaningof{\binpar{Q}{Q'}} \defneqls$
  
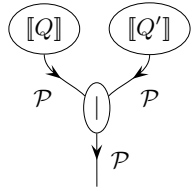
\[\begin{pspicture}(-1,0)(0,3.5)
    \rput(0,3){\ovalnode{A}{$\meaningof{Q}$}}
    \rput(2,3){\ovalnode{B}{$\meaningof{Q'}$}}
    \rput(1,1.5){\ovalnode{C}{$|$}}
    \pnode(1,0){D}
    \nccurve[angleB=135]{A}{C} \nbput{$\mathcal{P}$}
    \nccurve[angleB=45]{B}{C} \naput{$\mathcal{P}$}
    \nccurve{C}{D} \naput{$\mathcal{P}$}
  \end{pspicture}\]
\end{description}
For example, $\meaningof{(\mathsf{new}\;y)(\mathsf{new}\;x){x}{?}{( y_1, \ldots, y_n )} \Rightarrow {Q}}_{top}$ where $z$ is free in $Q$ is
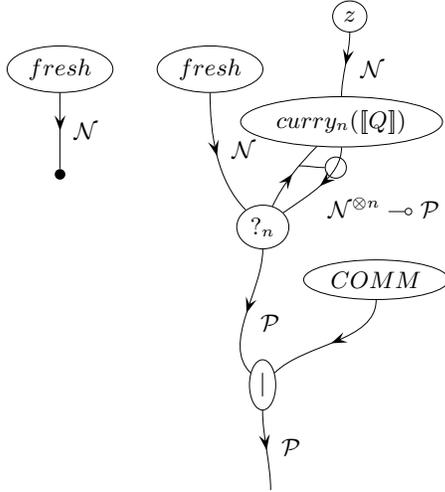
\begin{center}
  \begin{pspicture}(-2,-4)(2,5.5)
    \rput(1,2){\ovalnode{A}{$?_n$}}
    \rput(1,-1){\ovalnode{B}{$|$}}
    \nccurve[angleB=135]{A}{B} \naput{$\mathcal{P}$}
    \rput(0,5){\ovalnode{C}{$fresh$}}
    \nccurve[angleB=135]{C}{A} \naput{$\mathcal{N}$}
    \rput(2.5,4){\ovalnode{D}{$curry_n(\meaningof{Q})$}}
    \nccurve[angleA=65,angleB=-135]{A}{D} \ncput[npos=.65]{\pnode{clasp1}}
    \nccurve[angleA=-90,angleB=35]{D}{A} \naput{$\mathcal{N}^{\otimes n}\multimap \mathcal{P}$} \ncput[npos=.25]{\cnode{4pt}{clasp2}}
    \nccurve[angleA=0,angleB=180,ArrowInside=]{clasp1}{clasp2}
    \rput(3,1){\ovalnode{E}{$COMM$}}
    \nccurve[angleB=45]{E}{B}
    \pnode(1,-3){F}
    \nccurve{B}{F} \naput{$\mathcal{P}$}
    \rput(-3,5){\ovalnode{F}{$fresh$}}
    \rput(-3,3){\cnode*{2pt}{G}}
    \nccurve{F}{G} \naput{$\mathcal{N}$}
    \rput(2.5,6){\ovalnode{H}{$z$}}
    \nccurve{H}{D} \naput{$\mathcal{N}$}
    
  \end{pspicture}
\end{center}

\subsubsection{Bisimulation again}
In this setting we can provide a direct interpretation of observation
and bisimulation. Roughly speaking, $\meaningof{P}$ reduces to
$\meaningof{Q}$ just when we can apply the 2-morphism $comm_n$ to the
former to produce the latter. Since all non-trivial 2-morphisms are
\emph{generated} by $comm_n$, single step reductions $\meaningof{P}
\red \meaningof{Q}$ correspond precisely to the decomposion of a
$\meaningof{P}$ in terms of a ``context'' functor $C$, such that
$\meaningof{P}=C[src(comm_n)]$, and $\meaningof{Q} =
C[trgt(comm_n)]$. More generally, the interpretation of a term context
$\meaningof{K}$ is a functor from one hom category to another: the
functor takes an appropriate morphism $f$ to fill the hole and returns
a new morphism $\meaningof{K}( f )$; similarly, it takes a 2-morphism
$\alpha\maps f \Rightarrow f'$ between appropriate morphisms and
whiskers and/or tensors $\alpha$ with identity 2-morphisms to produce
a new 2-morphism $\meaningof{K}( \alpha )$.

\begin{defn}
  $\meaningof{P} \Rightarrow^{comm} \meaningof{P'}$ iff there is a
  2-cell, $F : \meaningof{P} \rightarrow \meaningof{P'}$ generated by
  exactly one top level occurrence of $comm_n$ and horizontal and
  vertical composition of identity 2-morphisms.
\end{defn}

\begin{lem}[reduction]
  $P \red P' \iff \meaningof{P} \Rightarrow^{comm} \meaningof{P'}$     
\end{lem}\label{redlemma}

\emph{Proof}: by construction. The only subtlety here is that there be
only one $COMM$ map to ensure only 1 component of $P$ reduces, but
this is just what the definition ensures. \hfill $\square$

Likewise, we can transport the notion of observability to the
categorical setting as a relation, $\Downarrow$, between 1-morphisms (not necessarily
in the same hom-category). More precisely, $\Downarrow$ is the smallest relation satisfying

\begin{itemize}
  \item $\meaningof{x!(y_1,\ldots,y_n)} \Downarrow \meaningof{x}$
  \item $\meaningof{P} \Downarrow \meaningof{x}$ or $\meaningof{Q} \Downarrow \meaningof{x}$ implies $\meaningof{P} \juxtap \meaningof{Q} \Downarrow \meaningof{x}$
  \item $\meaningof{P} \Downarrow \meaningof{x}$, $x \neq u$ implies $\meaningof{(\mathsf{new}\; u)P} \Downarrow \meaningof{x}$
\end{itemize}

\begin{lem}[observation]
  $P \downarrow x \iff \meaningof{P} \Downarrow \meaningof{x}$     
\end{lem}\label{oblemma}

\emph{Proof}: by construction. \hfill $\square$

Taken together these two notions provide an immediate lifting of the syntactic
notion of bisimulation to a corresponding semantic notion, which we write, $\wbbisim$.

\begin{defn}
  $\meaningof{P} \wbbisim \meaningof{Q}$ iff
  \begin{enumerate}
  \item If $\meaningof{P} \Rightarrow^{comm} \meaningof{P'}$ then $\meaningof{Q} \Rightarrow^{comm} \meaningof{Q'}$ and $\meaningof{P'} \wbbisim \meaningof{Q'}$.
  \item If $\meaningof{P} \Downarrow \meaningof{x}$, then $\meaningof{Q} \Downarrow \meaningof{x}$.
  \end{enumerate}
\end{defn}

\subsubsection{Full abstraction and contextual congruence}

\begin{thm}[full abstraction]
  $P \wbbisim Q \iff \meaningof{P} \wbbisimsem \meaningof{Q}$     
\end{thm}

\emph{Proof}: this follows from lemmas 4.1.2 and 4.1.3. \hfill $\square$

Typically, bisimulation is too rigid. Contextual congruence allows for
appropriate notion of equivalence in the presence of substitutions.

\begin{defn}[Contextual congruence]
  $P \simeq Q$ iff $C[P] \wbbisimsem C[Q]$ for all $C$.
\end{defn}

We need the corresponding notion

\begin{defn}[Contextual congruence]
  $\meaningof{P} \wcc \meaningof{Q}$ iff $\meaningof{C}( \meaningof{P} ) \wbbisim \meaningof{C}( \meaningof{Q} )$ for all $C$.
\end{defn}

where $\meaningof{C}$ is the
functor on hom categories mentioned above. We can immediately verify that

\begin{equation*}
  \meaningof{C}( \meaningof{P} ) = \meaningof{C[P]}
\end{equation*}

We require

\begin{equation*}
  P \simeq Q \iff \meaningof{P} \wcc \meaningof{Q}
\end{equation*}

But this follows directly

\begin{equation*}
  \begin{aligned}
    C[P] \wbbisimsem C[Q] \\
    \iff \mbox{(bisimilarity result)} \\
    \meaningof{C[P]} \wbbisimsem \meaningof{C[Q]} \\
    \iff \mbox{(definition of $\meaningof{C}$)} \\
    \meaningof{C}( \meaningof{P} ) \wbbisimsem \meaningof{C}( \meaningof{Q} )
  \end{aligned}
\end{equation*}

\section{Conclusions and future work}
We presented a fully abstract higher categorical semantics for the
{\pic}. Our semantics can be seen as a natural extension of
Curry-Howard in the categorical setting: if terms are taken to be
1-morphisms, then rewrites between terms should be 2-morphisms. Such
an approach is natural from another perspective in that it makes
comparison with operational semantics considerably simpler, at least
conceptually. To that end, we have already applied the approach to
models of other milestone computational calculi, such as the lazy
$\lambda$-calculus, with some initial success and hope to report on
that in subsequent papers. 

Perhaps more importantly, establishing connections like this between
two different computational frameworks should allow for transport of
other key conceptual tools. Here, we were able to transport a version
bisimulation to the categorical setting in a simple and
straightforward manner. It would be quite interesting to be able to
transport categorical notions of typing back to the process
setting. For example, Mellies and Zeilberger's refinement types
\cite{DBLP:conf/popl/MelliesZ15}, expressed as functors, suggest an
intriguing approach to a more categorical account of behavioral types.

Finally, the use of $COMM$ to control $comm_n$ based rewrites is
strongly reminiscent of the distinction between logical, or virtual
concurrency such as may be found in a threads package or operating
system process abstraction, versus actual hardware resources. Allowing
more than one $COMM$ resource provides, on the one hand a very natural
account of so-called true concurrency semantics, and on the other the
means to reason about these very practical situations which we hope to
investigate in future work.

\paragraph{Acknowledgments}
We would like to acknowledge Jamie Vicary for some early conversations
about a higher-category-based approach; Tom Hirschowitz for asking
some thoughtful and stimulating questions about earlier versions of
this work; and Marius Buliga for some initial conversations which
prompted us to reconsider enzymatic-style solutions.

\bibliographystyle{amsplain}
\bibliography{hctm}



\end{document}